\documentclass[rnote]{aa} 

\usepackage{graphicx}
\usepackage{txfonts}
\usepackage{mathrsfs}
\usepackage{hyperref}
\usepackage{breakurl}
\usepackage{subfigure}
\usepackage{color}
\usepackage{soul}
\def \xmm {\emph{XMM-Newton}}
\def \sax {\emph{BeppoSAX}}

\def \chandra {{\it Chandra}}

\def \src {\mbox{SGR~1900+14}}

\begin{document}

   \title{Ten years of INTEGRAL observations of the hard X-ray emission from SGR~1900+14}


   \author{L. Ducci
          \inst{1,2}
          \and 
          S. Mereghetti
          \inst{3}
          \and
          D. G\"otz
          \inst{4}
          \and
          A. Santangelo
          \inst{1}
          }

   \institute{Institut f\"ur Astronomie und Astrophysik, Eberhard Karls Universit\"at, 
              Sand 1, 72076 T\"ubingen, Germany\\
              \email{ducci@astro.uni-tuebingen.de}
              \and
              ISDC Data Center for Astrophysics, Universit\'e de Gen\`eve, 16 chemin d'\'Ecogia, 1290 Versoix, Switzerland
              \and
              INAF -- Istituto di Astrofisica Spaziale e Fisica Cosmica, Via E. Bassini 15, 20133 Milano, Italy
             \and
             AIM (UMR 7158 CEA/DSM-CNRS-Universit\'e Paris Diderot) Irfu/Service d'Astrophysique, Saclay, F-91191 Gif-sur-Yvette Cedex, France
             }

   \date{Received ...; accepted ...}

 
  \abstract
   {We exploited the high sensitivity of the INTEGRAL IBIS/ISGRI instrument to study
    the persistent hard X-ray emission of the soft gamma-ray repeater \src,
    based on $\sim 11.6$~Ms of archival data. 
    The 22$-$150~keV INTEGRAL spectrum can be well fit by a power law with photon index 
    $1.9 \pm 0.3$ and flux $F_{\rm x} = (1.11 \pm 0.17) \times 10^{-11}$~erg~cm$^{-2}$~s$^{-1}$ ($20-100$~keV). 
    A comparison with the 20$-$100~keV flux measured in 1997 with \sax, 
    and possibly associated with \src, shows a luminosity decrease by a factor of $\sim5$.
    The slope of the power law above 20~keV is consistent within the uncertainties with that
    of SGR~1806$-$20, the other persistent soft gamma-ray repeater for which a hard X-ray emission extending up
    to 150~keV has been reported.}


   \maketitle
%

\section{Introduction}

Soft Gamma-ray Repeaters (SGRs), together with Anomalous X-ray Pulsars (AXPs),
are high energy sources which are thought to be magnetars, i.e. neutron stars 
whose emission is powered mainly by their strong magnetic field
(see \citealt{Mereghetti15} for a recent review).

SGRs show recurrent short bursts (typical duration of the order
of $\approx 0.1$~s) in the X-ray and soft gamma-ray range, 
with peak luminosities of $10^{39}-10^{42}$~erg~s$^{-1}$
for normal bursts and up to $\approx 10^{47}$~erg~s$^{-1}$
for ``giant flares''.
Bursts are usually emitted during active states that can last
from a few days to months.
Active states are interrupted by long quiescent (or non-bursting)
time intervals with persistent luminosities of $\approx 10^{33}-10^{36}$~erg~s$^{-1}$.

In this paper we make use of the archival data 
of the INTErnational Gamma-Ray Astrophysics Laboratory (INTEGRAL, \citealt{Winkler03})
collected in the period 2003-2013 to study the hard X-ray emission (above 20~keV)
from SGR~1900+14.

Since the launch of INTEGRAL, SGR~1900+14 has shown only two periods of 
bursting activity: November 2002 \citep{Hurley02} and March 2006
(\citealt{Vetere06}; \citealt{Golenetskii06}).
Its persistent emission has been extensively studied,
especially in the soft X-ray range ($\lesssim 10$~keV)
with \xmm\ (e.g. \citealt{Mereghetti06}), \emph{Chandra} (e.g. \citealt{Gogus11}),
\emph{Suzaku} \citep{Nakagawa09}, and \emph{BeppoSAX} 
(\citealt{Tiengo07}; \citealt{Esposito07}).
The \xmm\ and \emph{Suzaku} spectra (0.8$-$10~keV) are well
fit either by an absorbed ($N_{\rm H}\approx 2.1\times 10^{22}$~cm$^{-2}$)
power law ($\Gamma \approx 1.9$ for \xmm, 
$\Gamma \approx 2.8$ for \emph{Suzaku}) plus a blackbody component
($kT \approx 0.5$~keV) or by two absorbed blackbodies ($kT_1 \approx 0.5$~keV,
$kT_2 \approx 1.9$~keV; \citealt{Mereghetti06}; \citealt{Nakagawa09}).
Persistent hard X-ray emission from \src\ in the $20-100$~keV
has been discovered by \citet{Gotz06} using observations 
obtained with INTEGRAL in the period 2003-2004.
The hard X-ray emission detected with INTEGRAL was fit 
by a power law with photon index $\Gamma \sim 3.1$,  
significantly steeper than that of other SGRs and AXPs \citep{Gotz06}.
\citet{Esposito07} reported the detection of hard ($20-150$~keV)
X-ray emission from a region around \src\ with the non-imaging spectrometer PDS
(Phoswich Detection System, \citealt{Frontera97}) on board the \emph{BeppoSAX}
satellite. They modeled the observed X-ray emission with 
a power law with slope $\Gamma \approx 1.6$.
Three transient X-ray sources were located within the PDS field of view,
but none of them was in a bright state during the observation of \src.
Therefore, the X-ray emission detected by \sax\ was likely produced by \src,
although the contamination from other unknown transient sources within the PDS
field of view cannot be ruled out \citep{Esposito07}.

The hard X-ray emission of \src\ has been
studied so far with a limited set of data. 
In this work, we exploit $\sim$10 years of archival INTEGRAL
data to further investigate the properties of  
its emission above 20~keV,
and put them in the broader context of the class of magnetars.
In Sect. \ref{sect. data analysis} we present the INTEGRAL observations
on which our work is based, the data analysis procedure, and the results,
that are then discussed in Sect. \ref{sect. discussion}.

\begin{figure}
\begin{center}
\includegraphics[bb=46 186 568 636,clip,width=\columnwidth]{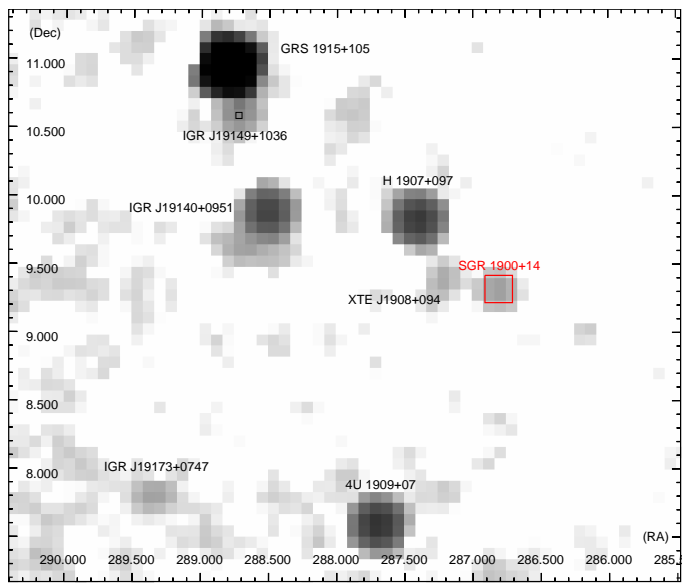}
\end{center}
\caption{IBIS/ISGRI mosaic image (in cts\,s$^{-1}$) 
of the \src\ field in the $22-50$~keV band.}
\label{mosaic}
\end{figure}

\section{Data analysis and results}
\label{sect. data analysis}

We used data obtained with the ISGRI (INTEGRAL Soft Gamma-Ray Imager,
\citealt{Lebrun03}) detector
of the coded-mask telescope IBIS (Imager on board
INTEGRAL Satellite, \citealt{Ubertini03}).
ISGRI operates in the $\sim15-400$~keV band.
IBIS has a fully coded field of view of $9^\circ \times 9^\circ$
and a partially coded field of view of $29^\circ \times 29^\circ$.

We performed the reduction and analysis of IBIS/ISGRI data
using the Off-line Science Analysis (OSA) 10.1 software 
provided by the ISDC Data Centre for Astrophysics
(\citealt{Goldwurm03}; \citealt{Courvoisier03}).
We analysed all the public data between 2003 March and 2013 June
where \src\ was within $12^\circ$ from the centre of the IBIS/ISGRI
field of view. At larger off-axis angle the IBIS response is not well 
known\footnote{See the INTEGRAL data analysis documentation: 
\url{http://www.isdc.unige.ch/integral/analysis}}.
We excluded all the data taken during bad time intervals.
The resulting data set consists of 4706 pointings,
corresponding to an exposure time of $\sim 11.6$~Ms.

Sky images of each pointing were generated in the energy band $22-50$~keV.
\src\ was never detected, being below the 5$\sigma$ threshold of 
detection, in the individual images.

We combined the individual images of the whole data set
to produce the total (mosaic) image (see Fig. \ref{mosaic}), 
in which \src\ is detected with a significance of 9.3$\sigma$.
The pixel significance distribution of a 
sky image obtained from the convolution of the 
detector image\footnote{The detector image is the shadowgram projected by the coded mask
of the telescope on the detector plane.} with a decoding array 
is expected to be Gaussian with zero mean and unitary standard deviation.
Systematic errors that are not taken into account by the algorithm that
reconstructs the sky image can lead to a broadening of this distribution 
\citep{Fenimore78}.
Systematic effects in the reconstructed image can be originated by
the imperfect knowledge of the real instrument and can particularly affect
observations of crowded regions of the Galaxy (e.g. \citealt{Krivonos10}).
We fitted the pixel significance distribution with a Gaussian with 
standard deviation of 1.18.
Therefore, the significance of detection is reduced to $9.3\sigma/1.18 \approx 7.9\sigma$
(for further details, see e.g. \citealt{Belanger06}; \citealt{Li11}).

\begin{figure}
\begin{center}
\includegraphics[bb=25 233 574 608,clip,width=\columnwidth]{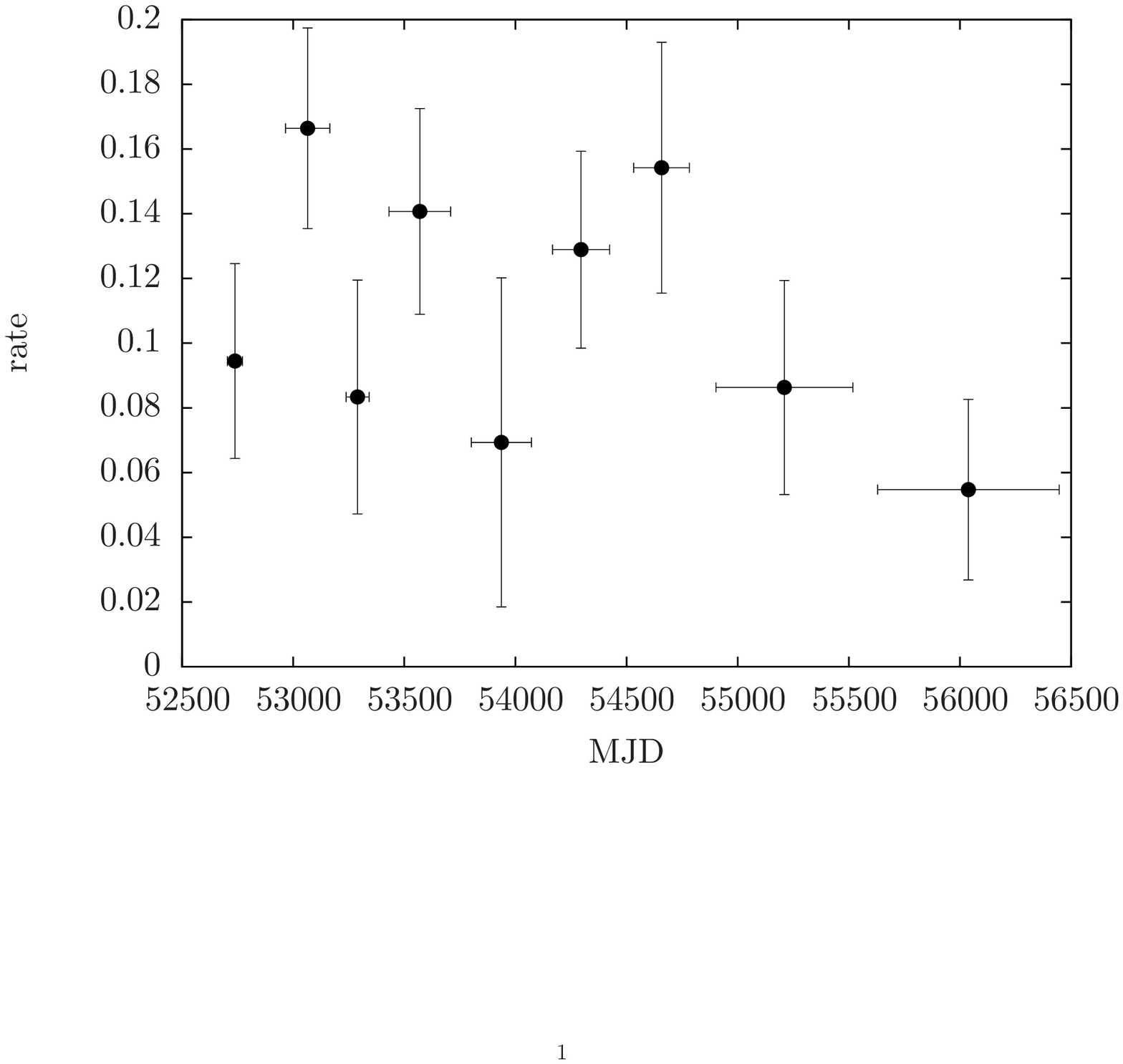}
\end{center}
\caption{$22-50$~keV IBIS/ISGRI lightcurve of \src. Time intervals and rates are reported in Table \ref{Table}.}
\label{lightcurve}
\end{figure}

\subsection{Lightcurve analysis}

We produced a lightcurve of \src\ (Fig. \ref{lightcurve})
extracting the fluxes from nine mosaic images corresponding 
to the time intervals indicated in Table \ref{Table}. Fitting these points 
with a constant, we obtain a reduced $\chi^2=1.48$ (8 d.o.f.).
Since this value has a large probability ($\sim 16$\%) of being obtained by a constant source, 
we conclude that there is no evidence for variability in the INTEGRAL data. 
Note, however, that the large error bars of the measured hard X-ray count rates 
do not allow us to reveal relatively small flux variations, 
such as those measured for \src\ at lower energy.
In fact, variations of $\sim 26$\% in the soft X-ray flux (0.8$-$6.5~keV)  
were detected with \xmm\ and \chandra\ during the period $\sim 52700-54520$~MJD 
that overlaps that of the INTEGRAL observations analysed in this work \citep{Gogus11}.  
Following \citet{Primini93}, we estimate that our observations could reveal 
at a $3\sigma$ confidence level only flux variation larger than $\sim$130\%.

\begin{table}
\begin{center}
\caption{Time intervals and intensities ($22-50$~keV)  
of the nine bins of the lightcurve of \src\ (Fig. \ref{lightcurve}). Errors are at $1\sigma$ confidence level.}
\label{Table}
\begin{tabular}{cccc}
\hline
\hline
Obs.   & T$_{\rm start}$ & T$_{\rm stop}$ &  cts~s$^{-1}$ \\
Number &   (MJD)   &   (MJD)   &              \\
\noalign{\smallskip}
\hline
\noalign{\smallskip}
1      & 52704.14  &  52772.10 &  0.09$\pm$0.03 \\
\noalign{\smallskip}
2      & 52965.66  &  53164.83 &  0.17$\pm$0.03 \\
\noalign{\smallskip}
3      & 53237.98  &  53341.66 &  0.08$\pm$0.04 \\
\noalign{\smallskip}
4      & 53431.43  &  53708.42 &  0.14$\pm$0.03 \\
\noalign{\smallskip}
5      & 53801.93  &  54072.02 &  0.07$\pm$0.05 \\
\noalign{\smallskip}
6      & 54166.75  &  54424.11 &  0.13$\pm$0.03 \\
\noalign{\smallskip}
7      & 54531.73  &  54783.05 &  0.15$\pm$0.04 \\
\noalign{\smallskip}
8      & 54902.04  &  55518.02 &  0.09$\pm$0.03 \\
\noalign{\smallskip}
9      & 55629.45  &  56446.46 &  0.06$\pm$0.03 \\
\noalign{\smallskip}
\hline
\end{tabular}
\end{center}
\end{table}

\subsection{Spectral analysis} \label{sect. spectral analysis}

We extracted the average spectrum of \src\ using the source
count rates obtained from the mosaic images in five energy bands 
(22$-$30, 30$-$38, 38$-$50, 50$-$70, 70$-$150~keV)
using the OSA tool {\tt mosaic\_spec}, which is particularly suited for faint sources.
We used the exposure-weighted average ancillary response file
and the rebinned response matrix specifically generated
for our particular data set.
Before fitting, we added systematic uncertainties of 5\% to the data set.
We obtained a good fit ($\chi^2_\nu=0.5$, 3 d.o.f.) with a power-law ($\Gamma=1.9 \pm 0.3$; 
uncertainties are at 90\% c.l.)
and $20-100$~keV flux $F_x = (1.11 \pm 0.17)\times 10^{-11}$~erg~cm$^{-2}$~s$^{-1}$.

\begin{figure}
\begin{center}
\includegraphics[bb=78 15 583 710,clip,angle=-90,width=\columnwidth]{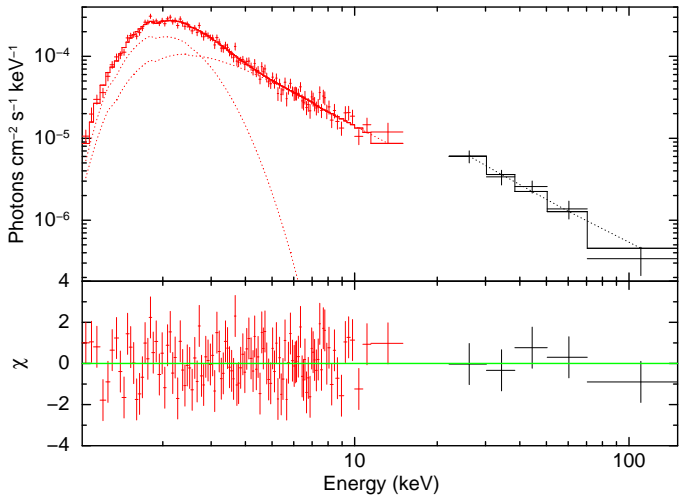}
\end{center}
\caption{Joint \xmm\ (red), IBIS/ISGRI (black) spectrum fitted with
an absorbed power law plus a blackbody model, with residuals
in units of standard deviations.}
\label{figure spec}
\end{figure}

Figure \ref{figure spec} shows the joint \xmm, IBIS/ISGRI spectrum.
We used the IBIS/ISGRI average spectrum obtained from the whole data set
and the \xmm\ 
data obtained on 2005 September 20 when the source was in quiescent state \citep{Mereghetti06}. 
The analysis of EPIC-pn data was performed with the 
\xmm\ Science analysis system (SAS) software, version 14.0.0.
We rejected time intervals affected by high background,
obtaining a total good exposure time of $\approx 20$~ks.
We included constant factors in the spectral fitting
to allow for normalization uncertainties between the instruments and 
differences between the soft and hard X-ray spectra due to 
source variability and not simultaneous observations.
We obtained acceptable fit ($\chi^2_\nu= 1.051$; 116 d.o.f.) 
of the joint \xmm, IBIS/ISGRI spectrum
with an absorbed ($N_{\rm H}=2.5\pm 0.2 \times 10^{22}$~cm$^{-2}$) 
blackbody ($kT=0.41\pm 0.04$~keV) plus a power-law component ($\Gamma=1.87\pm 0.17$).
These parameters are consistent with those derived 
by \citet{Mereghetti06} from the \xmm\ data alone.

\begin{figure*}
\begin{center}
\includegraphics[bb=22 229 563 610,clip,width=\columnwidth]{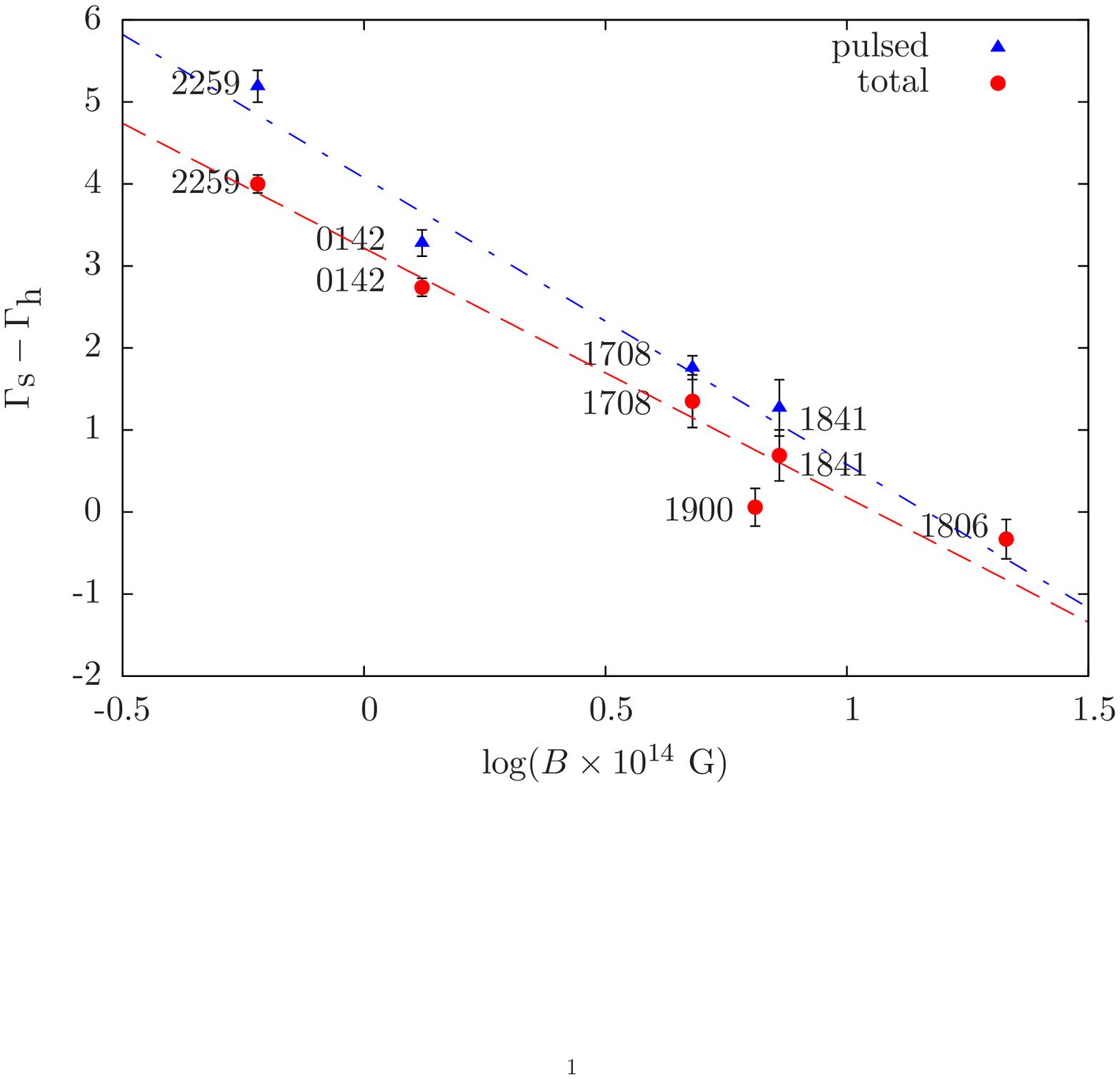}
\includegraphics[bb=22 229 563 610,clip,width=\columnwidth]{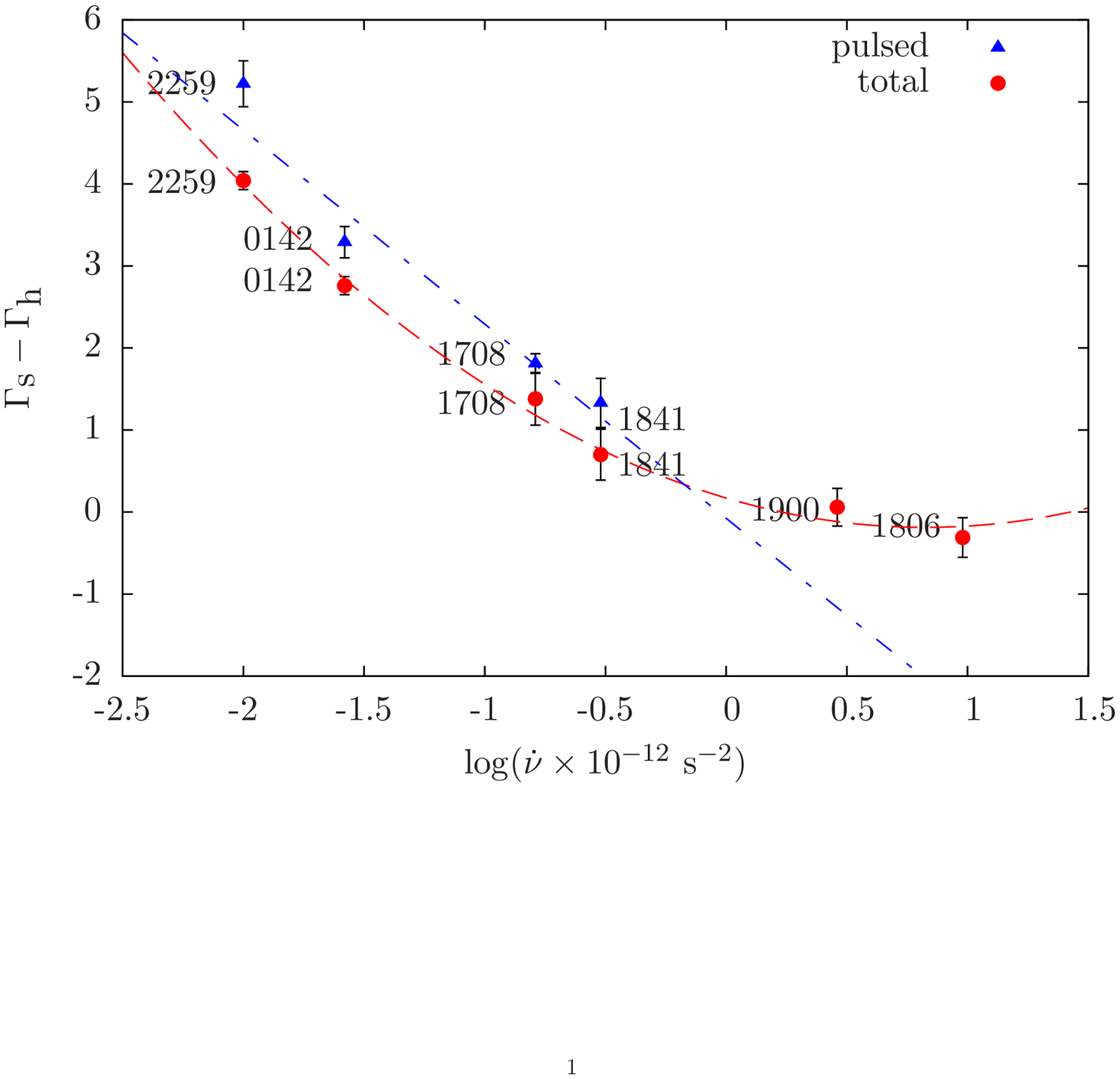}
\end{center}
\caption{Spectral turnover $\Gamma_{\rm s} - \Gamma_{\rm h}$ as function of the magnetic field $B$
         (left panel) and spin-down $\dot{\nu}$ (right panel) for all magnetars for which 
         $\Gamma_{\rm s}$ and $\Gamma_{\rm h}$ are measured. Red circles represent total flux and blue 
         triangles represent pulsed flux. Error bars represent $1\sigma$ uncertainties.
         Linear fits are shown on the left panel, both for pulsed flux (blue line) and total flux (red line),
         and on the right panel for pulsed flux (blue line). 
         For total flux data in the right panel, the best-fitting quadratic trend is shown.
         The figure is an updated version of figure 9 of \citet{Vogel14}, with the new value of 
         $\Gamma_{\rm s} - \Gamma_{\rm h}$ for \src\ obtained in this work.}
\label{figure kaspi}
\end{figure*}

\section{Discussion}
\label{sect. discussion}

We analysed 11.6~Ms of INTEGRAL data of \src, obtained from 2003 March to 2013 June.
We found that the $22-150$~keV spectrum is well fit by a power law with $\Gamma=1.9 \pm 0.3$
and flux $(1.11 \pm 0.17) \times 10^{-11}$~erg~cm$^{-2}$~s$^{-1}$.
A good agreement is found between the IBIS/ISGRI power law photon-index and the soft one, measured
below $\sim 10$~keV with \xmm\ during a quiescent state in 2005 \citep{Mereghetti06}.
The photon-index and significance of \src\ presented here are significantly different from
those obtained by \citet{Gotz06} ($\Gamma=3.1\pm0.5$, significance of 9$\sigma$)
which were based on INTEGRAL data collected in the period 
2003 March $-$ 2004 June (corresponding to Obs. number 1 of Table \ref{Table})
and analysed with OSA 5.1.
A comparison between the images obtained with OSA 10.1 and OSA 5.1 suggests that
the discrepancy in the results might be caused by the substantial improvement of the software
that reconstructs the images of the sky from the shadowgram projected
by the coded mask onto the detector plane.
The images processed with OSA 5.1 showed spurious features,
due to residual un-subtracted coding noise,
caused by the brightest source in the field
of view, GRS~1915+105, which contaminated the spectrum of \src.
Such features are not present anymore in the images reconstructed with OSA 10.1.
Therefore, the spectral results reported here supersede those of \citet{Gotz06}.

The slope of the hard X-ray spectrum derived with INTEGRAL 
is consistent with that measured with \sax\ before the giant flare of 1998, 
but the $20-100$~keV luminosity ($\approx 3 \times 10^{35}$~erg~s$^{-1}$ 
for $d=15$~kpc, \citealt{Vrba00}) is smaller by a factor $\sim 5$.
A similar decrease of the flux has been observed in the soft band, below $\sim 10$~keV
(\citealt{Mereghetti06}; \citealt{Gogus11}).

\citet{Kaspi10} noted an anti-correlation of the spectral turnover $\Gamma_{\rm s} - \Gamma_{\rm h}$
(where $\Gamma_{\rm s}$ is the photon-index in the soft band, $\lesssim 10$~keV,
and $\Gamma_{\rm h}$ is the photon-index in the hard band, $\gtrsim 10$~keV)
with inferred magnetic field $B$ and spin-down $\dot{\nu}$.
These authors gave a qualitative interpretation for the anti-correlation
$\Gamma_{\rm s} - \Gamma_{\rm h}$ vs $B$ in the framework of models in which the X-ray emission 
of magnetars strongly depends on magnetospheric currents. For magnetars with higher $B$,
$\Gamma_{\rm s}$ is harder because the scattering optical depth between the
relativistic electrons and the surface thermal photons increases.
This in turn has an effect on the population of electrons that do not scatter and instead
impact on the surface, heating it and creating a corona: fewer electrons
will impact on the surface, resulting in a softer $\Gamma_{\rm h}$.

Figure \ref{figure kaspi} shows an updated version of the plots $\Gamma_{\rm s} - \Gamma_{\rm h}$
vs $B$ and $\dot{\nu}$. Besides the magnetars considered by 
\citet{Kaspi10} and \citet{Vogel14} (1E~2259+568, 4U~0142+61, 1RXS~1708$-$40,
\src, 1E~1841$-$045, and SGR~1806$-$20),
we included our new result for \src\ and the pulsed flux of 1E~2259+568 from \citet{Vogel14}.
To quantify the significance of the anticorrelation, we determined the
Pearson's linear coefficient $r$ and the null hypothesis probability $p$. 
For the total flux data, we found $r=-0.97$, $p=0.0012$ 
and $r=-0.96$, $p=0.0025$, for $\Gamma_{\rm s} - \Gamma_{\rm h}$
as function of $B$ and $\dot{\nu}$, respectively.
For the pulsed data, we found $r=-0.98$, $p=0.0157$ and $r=-0.98$, $p=0.0236$
for $\Gamma_{\rm s} - \Gamma_{\rm h}$
as function of $B$ and $\dot{\nu}$, respectively.
The new value of the spectral turnovers of \src\ ($0.06 \pm 0.23$ compared to
the previous value $-1.2 \pm 0.5$) and 1E~2259+568 \citep{Vogel14}
improve the significance of the anticorrelation measured by \citet{Kaspi10}.
We used a linear model to fit the data 
($\Gamma_{\rm s} - \Gamma_{\rm h}=a + b\log_{10}x$, where $x=B$ or $x=\dot{\nu}$).
For $\Gamma_{\rm s} - \Gamma_{\rm h}$ as function of $B$ and total flux data set
we found $a=46\pm2$, $b=-3.04\pm0.14$, $\chi^2=15.9$ (4 d.o.f.).
For the pulsed flux we obtained $a=53\pm3$, $b=-3.5\pm0.2$, $\chi^2=9.2$ (2 d.o.f.).
When considering $\Gamma_{\rm s} - \Gamma_{\rm h}$ as function of $\dot{\nu}$,
we found for total flux: $a=-17.4\pm 0.9$, $b=-1.51\pm0.07$, $\chi^2=29.9$ (4 d.o.f.);
for the pulsed flux: $a=-29 \pm 3$, $b= -2.4\pm 0.2$, $\chi^2=8.3$, (2 d.o.f.).
A quadratic model ($\Gamma_{\rm s} - \Gamma_{\rm h}=a + b\log_{10}\dot{\nu} + c(\log_{10}\dot{\nu})^2$)
fits better the total flux data: $a=65\pm 15$, $b=12\pm 2$, $c=0.5\pm 0.1$, $\chi^2=2.1$ (3 d.o.f.).
Best fit models are shown in Figure \ref{figure kaspi}.

\section{Conclusions}

We have reported a new analysis of the hard X-ray emission from 
\src\ based on ten years of observations with the IBIS/ISGRI
instrument on the INTEGRAL satellite. Due to the improved
calibrations and analysis software, we obtained results which
differ from and supersede those reported earlier for this source and
based only on the first two years of data.
We measured a 20$-$100~keV flux of $F_{\rm x} = (1.11 \pm 0.17) \times 10^{-11}$~erg~cm$^{-2}$~s$^{-1}$,
which is lower than that seen in 1997 with \sax\ by a factor of $\sim 5$.
While the previously found photon index at $E>20$~keV indicated
a power law spectrum significantly softer than those of other 
SGRs and AXPs, the new value of $1.9 \pm 0.3$ is similar to those 
found for the other sources and to that measured at lower energy
for \src.

\begin{acknowledgements}

This work is partially supported by the Bundesministerium f\"ur
Wirtschaft und Technologie through the Deutsches Zentrum f\"ur Luft
und Raumfahrt (grant FKZ 50 OG 1301).
This work has been partially supported through financial contribution 
from the agreement ASI/INAF I/037/12/0 and from PRIN INAF 2014.
This paper is based on data from observations with INTEGRAL, and \xmm.
INTEGRAL is an ESA project with instruments and science data centre funded by ESA
member states (especially the PI countries: Denmark, France, Germany,
Italy, Spain, and Switzerland), Czech Republic and Poland,
and with the participation of Russia and the USA.
\xmm\ is an ESA science mission with instruments and contributions directly
funded by ESA Member States and NASA.
This research has made use of data and/or software provided by the High Energy Astrophysics Science Archive Research Center (HEASARC), which is a service of the Astrophysics Science Division at NASA/GSFC and the High Energy Astrophysics Division of the Smithsonian Astrophysical Observatory.
This research has made use of SAOImage DS9, 
developed by Smithsonian Astrophysical Observatory.
\end{acknowledgements}

\bibliographystyle{aa} 
\bibliography{lducci_sgr1900}

\end{document}